\documentclass[12pt,preprint]{aastex}
\usepackage{lineno}
\usepackage{ulem}
\usepackage{longtable}
\usepackage{array} 
\usepackage[fleqn]{amsmath}

\shorttitle{}
\shortauthors{}

\begin{document}

\received{}
\accepted{}

\title{A weak modulation effect detected in the light curves of KIC 5950759: intrinsic or instrumental effect?}

\author{Taozhi Yang, A. Esamdin$^{\star}$, Fangfang Song, Hubiao Niu, Guojie Feng, Peng Zong, Xiangyun Zeng, Junhui Liu, Jinzhong Liu, Lu Ma, Fei Zhao}
\affil{Xinjiang Astronomical Observatory, Chinese Academy of Sciences, Urumqi 830011, Xinjiang, China;  
\\yangtaozhi@xao.ac.cn, aliyi@xao.ac.cn}

\author{Taozhi Yang, A. Esamdin, Fangfang Song, Peng Zong, Xiangyun Zeng, Junhui Liu}
\affil{University of Chinese Academy of Sciences, Beijing 100049, China;}

\slugcomment{23/08/2018}


\begin{abstract}

In this paper, the high-precision light curves of the $Kepler$ target KIC 5950759 are analyzed. The Fourier analysis of the long cadence (LC) light curve reveals three independent frequencies. Two of them are main pulsation modes: F0 = 14.221373(21) $\rm{d^{-1}}$ and F1 = 18.337249(44) $\rm{d^{-1}}$. The third independent frequency, $f_m$ = 0.3193 d$^{-1}$, is found in LC data with a signal-to-noise ratio of 6.2. A weak modulation of $f_m$ to F0 and F1 modes (triplet structures centered on F0 and F1) are detected both in long and short cadence data. This is the first detection of the modulation effect in a double-mode high-amplitude $\delta$ Scuti (HADS) star. The most possible cause of the modulation effect in the light curves is amplitude modulation with the star's rotation frequency of 0.3193 d$^{-1}$. The preliminary analysis suggests that KIC 5950759 is in the bottom of the HADS instability strip and likely situated in the main sequence. Spectroscopic observations are necessary to verify the true nature of the modulation terms. 

\end{abstract}




\section{INTRODUCTION}   

The $Kepler$ $Space$ $Telescope$ was launched in 2009 March, and its main scientific goal was to search for the terrestrial planets using the transit method \citep{Borucki2010}. The asteroseismology program is an extremely important byproduct of the $Kepler$ mission \citep{Gilliland2010}. Due to the unprecedented photometric precision on the order of a few $\mu$mag with a high duty-cycle, and continuous collection of data over about 4 years, $Kepler$ data allows one to study the internal structure and physical process of stars with an ultra-high precision than that obtained with any telescope on earth \citep{Koch2010}. The $Kepler$ telescope is in a 372.5 day Earth-trailing orbit and its field of view covers 105 deg$^2$ in the constellations of Cygnus and Lyra. $Kepler$ provides two observation strategies for its targets, i.e. short cadence (SC) and long cadence (LC) with sampling times of 58.85 s and 29.424 minutes, respectively. The SC observations are adopted for the primary mission as they are able to obtain more data points during a candidate object passing in front of its host star. However, most of the targets are observed with LC in order to maximize the target's number through long exposure time. Consequently, more than 170,000 targets are available in LC observations \citep{Jenkins2010}. More than 2000 $\delta$ Sct stars have been detected through the $Kepler$ mission \citep{Balona2011,Balona2014,Bowman2016}. One of them, KIC 5950759, has been classified as a double-mode high-amplitude $\delta$ Sct (HADS) star by analyzing the LC data \citep{Bowman2016}.

The $\delta$ Sct stars are typical A- and F-type variable stars with period from 0.02 to 0.25 days. They are usually on or above the main sequence on the H-R diagram, and are situated in the lower classical instability strip \citep{Breger2000}. Their oscillations are mainly driven by the $\kappa$ mechanism that occurrs in the partial ionization zone of He $\rm{\uppercase\expandafter{\romannumeral2}}$ \citep{Gautschy1995,Breger2000}. The $\delta$ Sct stars can oscillate in both pressure and gravity modes \citep{Breger1995,Breger2000,Balona2016}. As a subclass of $\delta$ Sct stars, the HADS stars usually pulsate with a light amplitude larger than 0.3 mag and generally rotate slowly with $v$sin$i \leqslant 30 $ km s$^{-1}$ \citep{Breger2000}. Compared with the low-amplitude $\delta$ Sct stars, the HADS stars possess a more restrictive instability strip with a width in temperature of about 300 K and tend to shift to a lower temperature with evolution \citep{McNamara2000}. \cite{Lee2008} reveal that only about 0.24 percent of the stars suited in the $\delta$ Sct region belong to HADS stars. The majority of HADS stars are typically young and metal-rich Population $\rm{\uppercase\expandafter{\romannumeral1}}$ stars; some have been confirmed to be SX Phe variables, and are Population $\rm{\uppercase\expandafter{\romannumeral2}}$ metal-deficient stars \citep{Breger2000,Balona2012}. In general, the HADS stars pulsate with only one or two modes (e.g., AE UMa, \citealt{Niu2017}; YZ Boo, \citealt{Yang2018}; etc.), and most of their pulsations belong to radial modes. AI Vel \citep{Walraven1992} and V974 Oph \citep{Poretti2003} are both detected to have a number of millimag amplitude radial and nonradial modes although they are identified as radial mode variables. Indeed, some theoretical studies suggest that rotation effects can offer more additional constraints on the mode identification, especially for the nonradial modes rotationally coupled with the fundamental and first overtone radial modes \citep{Suarez2006a,Suarez2007}. Therefore, the analysis of the low-amplitude radial and nonradial modes in HADS stars may provide more and pivotal information about the internal structure of the stars. In this decade, with the launch of more space telescopes, e.g. $MOST$ \citep{Walker2003}, $CoRoT$ \citep{Baglin2006}, and $Kepler$ \citep{Borucki2010}, more pulsation modes and the long-term variations are also detected for HADS stars due to the continuous and ultra-high-precision time series \citep{Poretti2011,Balona2012a}.

KIC 5950759 ($\alpha_{2000}$= $19^{h}$$15^{m}$$00^{s}$.54, $\delta_{2000}$= +$41^{\circ}$$13^{'}$$55^{''}$.4) is thought to be a double-mode HADS star according to its period ratio of the two independent modes by \citet{Bowman2016}. \citet{Bowman2017} note that its pulsation periods of the two modes were increasing during the 4 years of observation. The Kepler magnitude of the star is Kp = 13.96 mag \citep{Huber2014}, and its $J$ = 13.293 (0.022) mag, $H$ = 13.129 (0.028) mag, and $K$ = 13.187 (0.036) mag in the 2MASS All Sky Catalog of Point Sources \citep{Cutri2003}. The SDSS photometry provides its $g$ = 14.056 mag, $r$ = 13.934 mag, $i$ = 14.013 mag, $z$ = 14.074 mag, and D51 = 13.995 mag \citep{Latham2005}. The star is also observed in the All Sky Automated Survey \citep{Pigulski2009} and the magnitudes: $V$ = 13.516 mag, $I$ = 12.950 mag, $\Delta$$V$ = 0.36 mag, and $\Delta$$I$ = 0.13 mag. The effective temperature of the star is 7840 $\pm$ 300 K in the Kepler Input Catalog \citep[KIC;][]{Brown2011}. After revising the stellar parameters of all of the Kepler targets, \cite{Huber2014} presents effective temperature of the star as $T_{\rm{eff}}$ = 8040 $\pm$ 270 K. The mass and radius of the star are derived as $M$ = $1.78^{+0.30}_{-0.26}$ $M_{\odot}$ and $R$ = $2.08^{+1.06}_{-0.41}$ $R_{\odot}$ (NASA Exoplanet Archive\footnote{NASA Exoplanet Archive: {https://exoplanetarchive.ipac.caltech.edu/index.html}}). There are no spectroscopic observations so far to verify the star's effective temperature and metallicity, and no new simultaneous multicolor observations to derive its mode identification.

\citet{Bowman2016} investigated the 12 strongest frequencies in the range 4 $\leqslant$ $\nu$ $\leqslant$ 24 d$^{-1}$ using the LC data. This paper presents an in-depth investigation on the light variations of the star through a wider frequency range using both LC and SC data.

\section{OBSERVATION AND DATA REDUCTION}

\begin{figure*}
\begin{center}
  \includegraphics[width=0.85\textwidth]{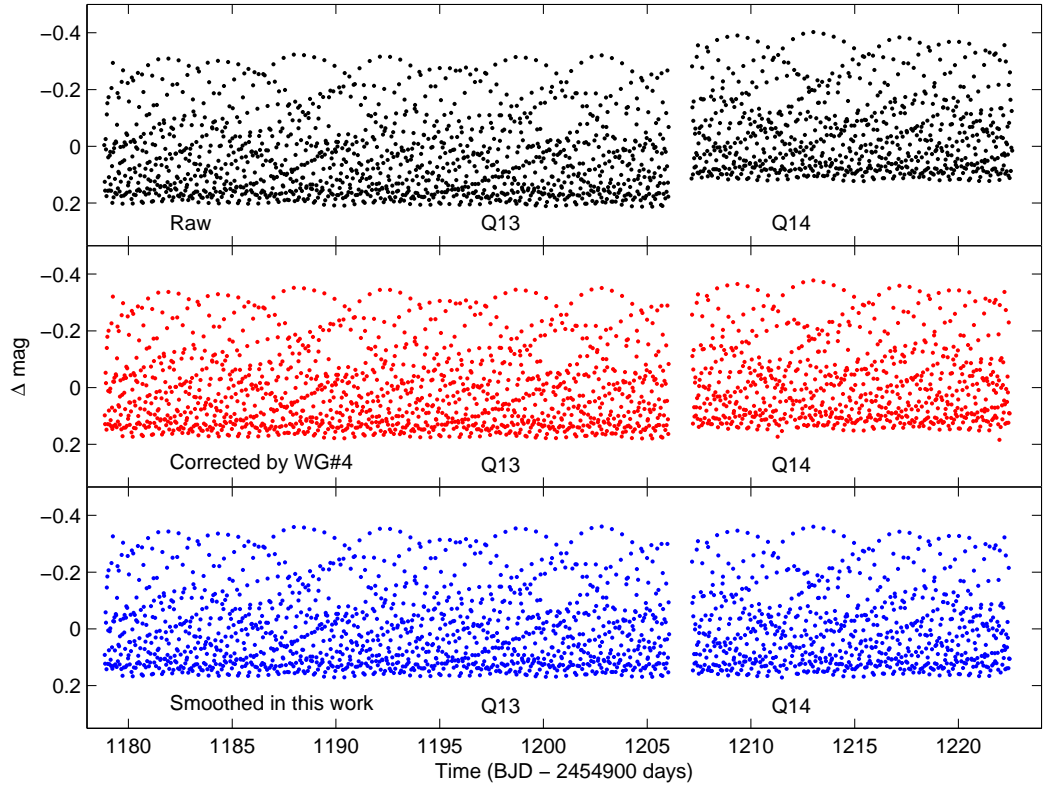}
  \caption{LC light curves of KIC 5950759. Top panel: the raw light curve shows a clear jump between different quarters; middle panel: the corrected light curve by WG$\#$4; bottom panel: the smoothed light curve in this work.}
    \label{fig:before_and_after}
\end{center}
\end{figure*}

\begin{figure*}
\begin{center}
  \includegraphics[width=0.90\textwidth]{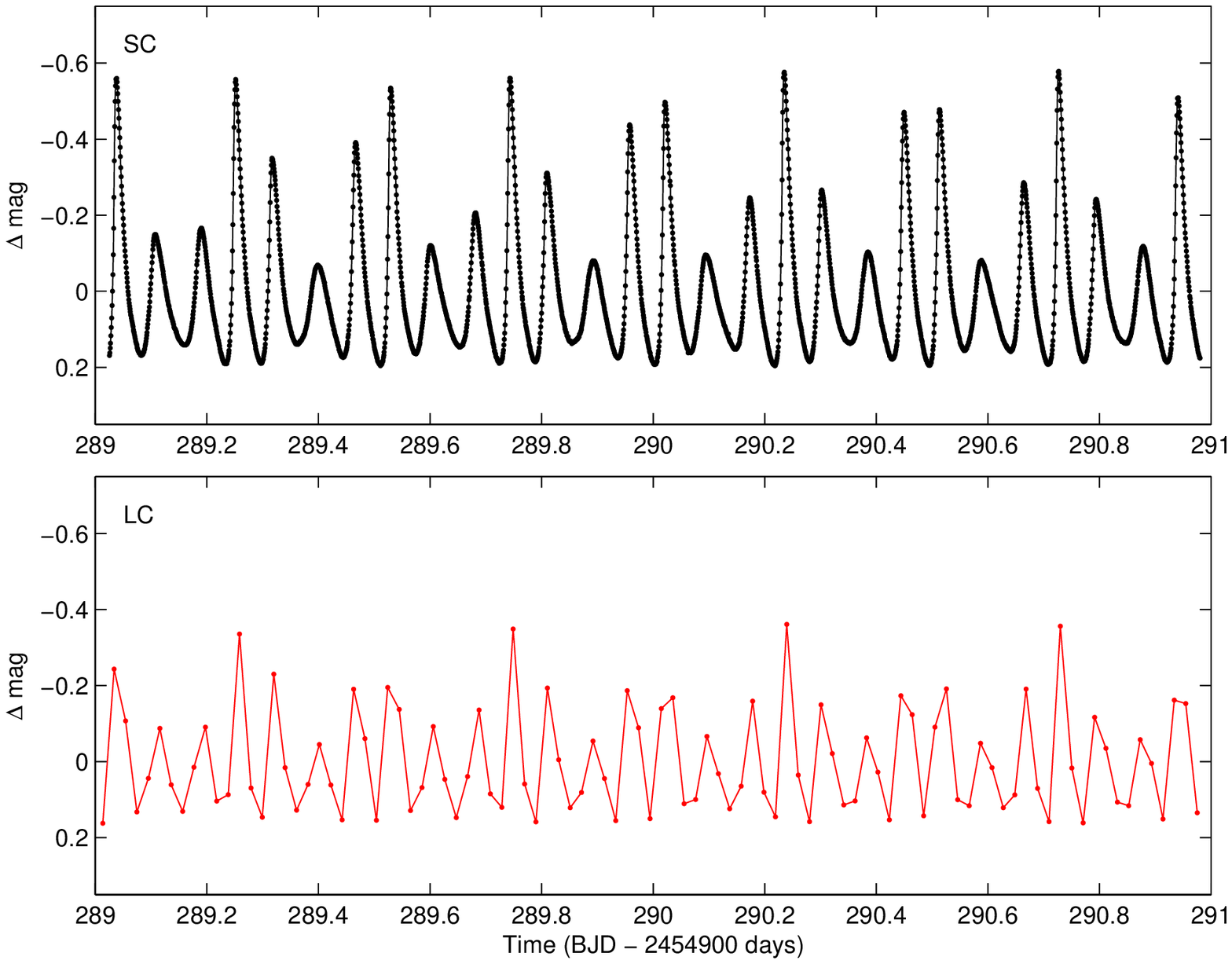}
  \caption{Sample of the SC and LC light curves of KIC 5950759. Top panel: light curve from the SC data. Bottom panel: light curve from the LC data. The amplitude of light curve from the LC data is lower than that from the SC data since the longer integration time (29.4 minutes) in the LC data heavily suppresses the amplitude of the LC light curve.}
    \label{fig:SC_LC}
\end{center}
\end{figure*}

KIC 5950759 was observed continuously from BJD 2455185.36 to 2455216.42 (Q4.1: 45449 SC observations) spanning 31 days, and from BJD 2454964.51 to 2456424.00 (Q1 - Q17: 64782 LC observations) spanning about 1460 days. Detailed characteristics of both $Kepler$ SC and LC data can be found in \citet{Gilliland2010} and \citet{Jenkins2010}, respectively. Both the SC and LC photometric time series data for KIC 5950759 are available through the Kepler Asteroseismic Science Operations Center (KASOC) data base\footnote{KASOC data base: {http://kasoc.phys.au.dk}}\citep{Kjeldsen2010} in two types: one is a type of flux labeled as "raw", which has been reduced by the NASA Kepler Science pipeline, and the other is corrected flux, which is made by KASOC Working Group 4 (WG$\#$4: $\delta$ Scuti targets). The second type has been corrected for cooling down, cooling up, outliers, and jumps. We use the corrected flux and convert it to the magnitude, then perform corrections eliminating the outliers and possible linear trends in some quarters of the LC data. The mean value of every quarter is then subtracted, and the smoothed time series is obtained. Figure \ref{fig:before_and_after} shows parts of the raw light curve (upper panel), the light curve corrected by WG$\#$4 (middle panel), and smoothed light curve in this work (bottom panel), respectively. Figure \ref{fig:SC_LC} shows a sample of SC and LC light curves with the same time period. The light amplitude of KIC 5950759 is about 0.8 mag in the SC data, and the amplitude of the LC light curve is lower than that of the SC since the longer integration time (29.4 minutes) in the LC data heavily suppresses the amplitude of the LC light curve.

\section{FREQUENCY ANALYSIS}

A Fourier analysis is performed for the smoothed LC and SC data of KIC 5950759 using PERIOD04 \citep{Lenz2005}, respectively. The light curves are fitted using the formula

\begin{equation}
m = m_{0} + \Sigma\mathnormal{A}_{i}sin(2\pi(\mathnormal{f}_{i}\mathnormal{t} + \phi_{i})), \label{equation1}
\end{equation}

where $m_{0}$ is the zero-point, $A_{i}$ is the amplitude, $f_{i}$ is the frequency, and $\phi_{i}$ is the corresponding phase.

In \citet{Bowman2016}, amplitude variation analysis is performed for all of the stars with frequency in the range of 4 $\leqslant$ $\nu$ $\leqslant$ 24 d$^{-1}$. The reasons for choosing the above frequency range are that typical pulsating frequency of $\delta$ Scuti stars is larger than 4 d$^{-1}$ and the Nyquist frequency of LC data is $f_{N}$ = 24.469 d$^{-1}$ \citep{Murphy2013a,Holdsworth2014}. In order to explore more potential frequencies, we search significant peaks in the frequency range of 0 $<$ $\nu$ $<$ 24.469 d$^{-1}$, which is wider than that of \citet{Bowman2016}. At each step in the process of extracting frequency, the highest peak was selected as a significant frequency, then a multiperiod least-squares fit using the Equation (\ref{equation1}) was performed to the data with all of the significant frequencies detected, and the solutions of them were obtained. A light curve constructed using the solutions was subtracted from the data, and the residual was obtained to search for significant frequency in the next step. The above steps were repeated until the first 12 frequencies were detected. And then we fixed the frequencies, amplitudes, and phases of these 12 frequencies for further searching, because all of these parameters have strong correlation with each other that will lead to nonconvergence \citep{John2011}. Then, the above steps were repeated until there was no significant peak in the residual. The criterion (signal-to-noise ratio (S/N) $>$ 4.0) suggested by \citet{Breger1993} was adopted to judge the significant peaks. The uncertainties of the frequencies were calculated following \citet{Kallinger2008}. 

\subsection{Frequencies in the LC data}

\begin{figure}
\begin{center}
  \includegraphics[width=0.9\textwidth]{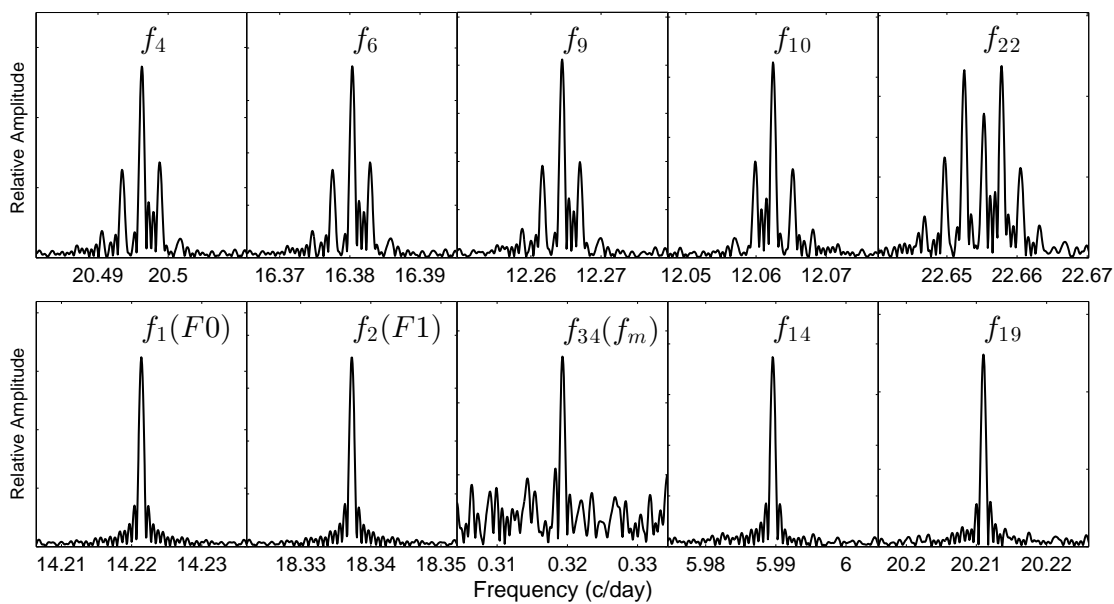}
  \caption{Typical spectra of the alias and real frequencies. The upper panel shows five alias frequencies ($f_{4}$, $f_{6}$, $f_{9}$, $f_{10}$, and $f_{22}$), which are multiplets with a frequency interval of $f_{\rm{orb}}$ = 0.00268 $\rm{d^{-1}}$; the bottom panel shows five real frequencies, including three independent frequencies ($f_{1}$, $f_{2}$, and $f_{34}$) and two combinations ($f_{14}$ and $f_{19}$).}
    \label{fig:real_alias_frequency}
\end{center}
\end{figure}

A total of 35 significant frequencies are detected in the LC data ($f_{1}$ to $f_{35}$ listed in Table \ref{tab:Frequency-LC}), including the 12 strong frequencies which are consistent with that in \citet{Bowman2016} and 23 new frequencies detected in this work. All of the alias frequencies and combinations confirmed are marked in the comment column of Table \ref{tab:Frequency-LC}. As noted by \citet{Bowman2016}, the alias frequency in the LC data can be easily identified due to the multiplet structure of its spectra peak, which is split by the orbital motion of the $Kepler$ spacecraft (the orbital frequency $f_{\rm{orb}}$ = 0.00268 d$^{-1}$). Upper panel in Figure \ref{fig:real_alias_frequency} shows the spectral sample of five alias frequencies ($f_{4}$, $f_{6}$, $f_{9}$, $f_{10}$, and $f_{22}$), bottom panel shows the spectra of five real frequencies ($f_{1}$, $f_{2}$, $f_{34}$, $f_{14}$, and $f_{19}$). Comparing with the real frequency, the multiplet structure of the alias frequency is obvious. More detail of the alias frequencies of the $Kepler$ data can be found in \citet{Murphy2013a}. For the linear combinations, we firstly find three frequencies with relationships like $f_a$ $\pm$ $\sigma_a$ $\approx$ ($f_b$ $\pm$ $\sigma_b$) $\pm$ ($f_c$ $\pm$ $\sigma_c$). Then we determine possible linear combinations only when those frequencies satisfy the relationship $\sigma_a$ $\leq$ 3 $\times$ ($\sigma_b$ + $\sigma_c$) as suggested by \citet{Fu2013}. 

As listed in Table \ref{tab:Frequency-LC}, three independent frequencies (i.e. $f_1$, $f_2$, and $f_{34}$) are detected in the LC data. $f_1$ and $f_2$ are the fundamental mode (F0) and the first overtone mode (F1) frequencies. The modulation of $f_{34}$ to $f_1$ and $f_2$ are clearly detected (i.e. $f_{15}$, $f_{28}$ and $f_{32}$, $f_{35}$, respectively). We mark $f_{34}$ as modulation frequency $f_{m}$ in Table \ref{tab:Frequency-LC}.

\subsection{Frequencies in the SC data}

\begin{figure}
\begin{center}
  \includegraphics[width=0.85\textwidth]{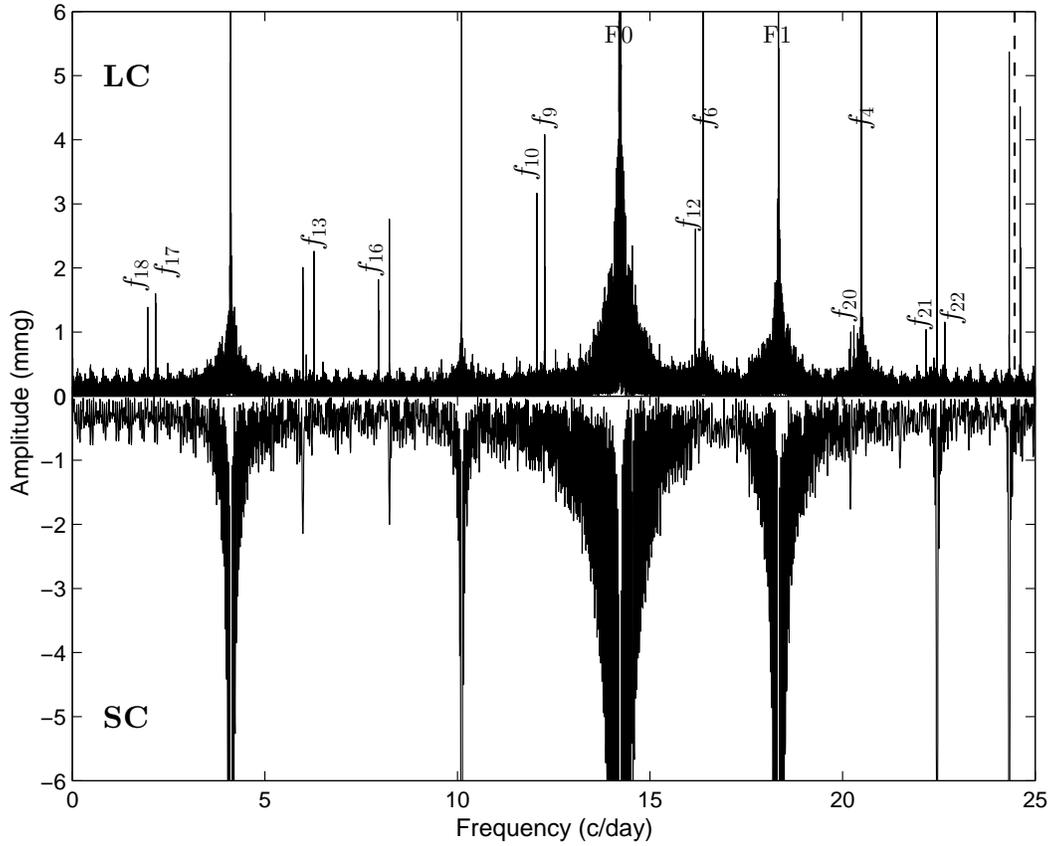}
  \caption{Amplitude spectra from LC (upper panel) and SC (bottom panel) data of KIC 5950759. Note that the maximum amplitude is limited to 6 mmg for showing the low-amplitude alias frequencies in the LC data. Twelve alias frequencies in the LC data are marked. The Nyquist frequency of the LC data is indicated by the vertical dashed line.}
    \label{fig:spectrum}
\end{center}
\end{figure}

For the SC data, although the Nyquist frequency of the SC data is $f_{\rm{SN}}$ = 734.1 d$^{-1}$, we choose a frequency range of 0 $<$ $\nu$ $<$ 50 d$^{-1}$, which covers the typical pulsation frequency of a $\delta$ Sct star. Table \ref{tab:Frequency-SC} lists all of the significant frequencies ($f_{S1}$ to $f_{S29}$) detected in the SC data, including 2 independent frequencies ($f_{S1}$ = F0 and $f_{S2}$ = F1), 3 harmonics ($f_{S3}$, $f_{S7}$ and $f_{S9}$), 13 combinations, and 11 modulation terms that are modulated by $f_{m}$. In the SC spectrum, seven frequencies ($f_{S1}$, $f_{S2}$ $f_{S5}$, $f_{S8}$, $f_{S10}$, $f_{S11}$, and $f_{S15}$) detected are consistent with that in \citet{Bowman2016} within 3$\sigma$ frequency accuracy, however, the other five frequencies in \citet{Bowman2016} do not appear in the SC spectrum because they are alias frequencies of the LC. High sampling frequency (1468.1 d$^{-1}$) of the SC data allows us to verify again the alias frequencies of the LC data using the significant frequencies of the SC data because the Nyquist alias of LC do not appear in SC spectrum. Figure \ref{fig:spectrum} compares the amplitude spectra of LC and SC data, and clearly shows that the alias frequencies (e.g. $f_{4}$, $f_{6}$, $f_{9}$, $f_{10}$, $f_{12}$, $f_{13}$, $f_{16}$, $f_{17}$, $f_{18}$, $f_{20}$, $f_{21}$, and $f_{22}$) in the spectrum of LC do not appear in that of SC. 

\begin{figure}
\begin{center}
  \includegraphics[width=0.85\textwidth]{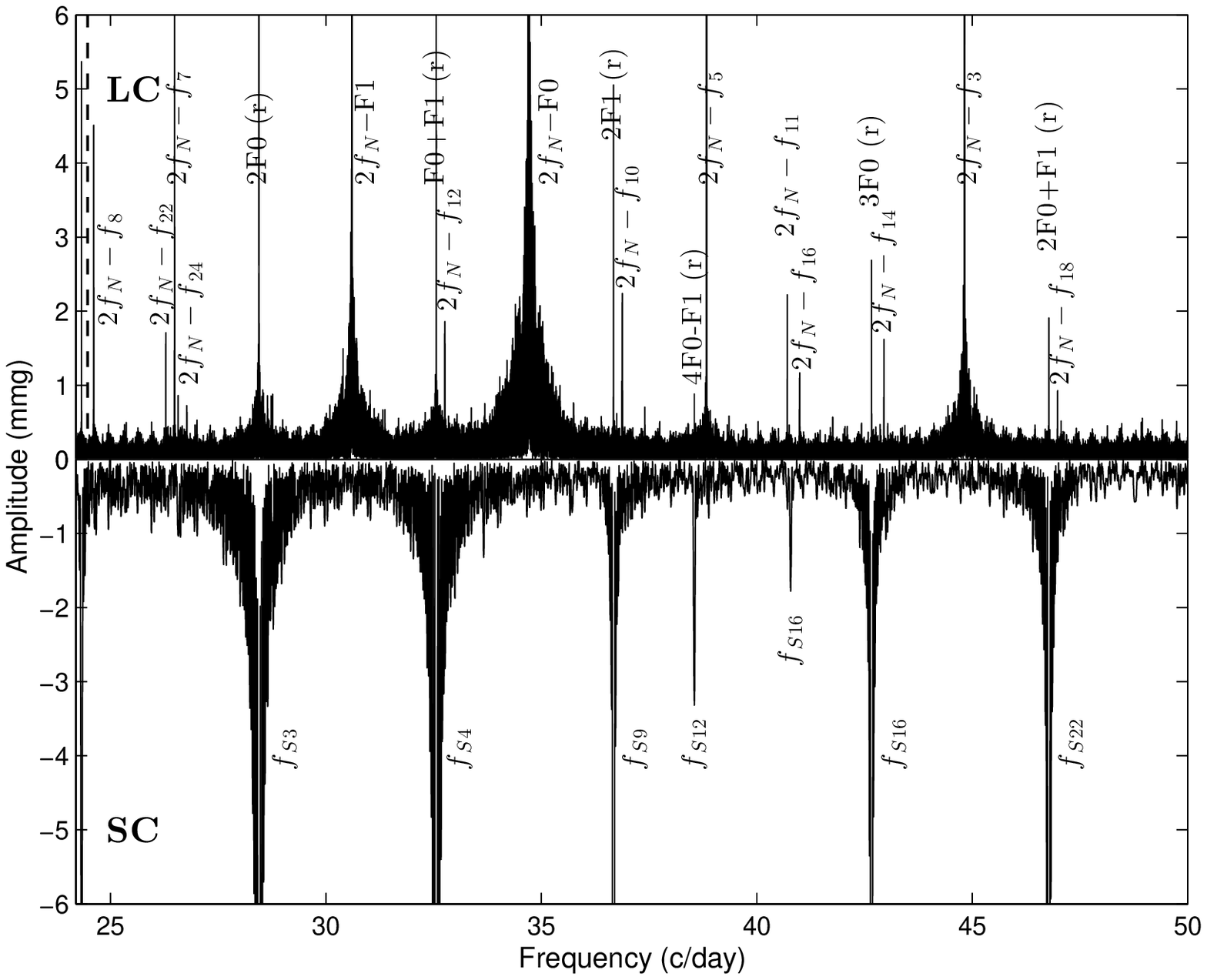}
  \caption{Amplitude spectra from LC (upper panel) and SC (bottom panel) data beyond the LC Nyquist frequency ($f_{N}$, the vertical dashed line). In the frequency range 24.469 $<$ $\nu$ $<$ 50 d$^{-1}$, the real frequencies beyond $f_{N}$ in LC spectrum are identified and labeled with (r). All the alias frequencies in LC spectrum are involved with $f_{N}$ and labeled in the corresponding position.}
    \label{fig:spectrum_half_last}
\end{center}
\end{figure}

We also show the amplitude spectra of the SC and LC data up to 50 d$^{-1}$ in Figure \ref{fig:spectrum_half_last}. In this frequency range (24.469 $<$ $\nu$ $<$ 50 d$^{-1}$), there are real frequencies (i.e. 2F0, F0+F1, 2F1, 4F0-F1, 3F0, and 2F0+F1; see the upper panel of Figure \ref{fig:spectrum_half_last}) in the LC spectrum, and they have corresponding frequencies in the SC spectrum (see the bottom panel of Figure \ref{fig:spectrum_half_last}). The other frequencies in the LC spectrum in this range are Nyquist aliases, they are combinations of LC Nyquist frequency ($f_{N}$) and real (or alias) frequencies in the range of 0 $<$ $\nu$ $<$ 24.469 d$^{-1}$. Their identifications have been labeled in corresponding positions in the upper panel of Figure \ref{fig:spectrum_half_last}. These alias frequencies do not appear in the SC spectrum.

As shown in Figure \ref{fig:real_alias_frequency}, $f_{m}$ is not alias frequency of LC data, we should note that $f_{m}$ is detected in LC but not in SC data, mainly because the noise in LC spectrum is lower than that in SC due to the data precision of LC observations is much higher than that of SC as a result of the much longer integration time applied for LC. The S/N of $f_{m}$ in LC spectrum is only 6.2, so that in SC spectrum is most possibly lower than our frequency detection threshold.

\begin{table*}
\begin{center}
\caption{All of the Frequencies Detected in the LC Data (Denoted by $f_i$)}
  \begin{tabular}{@{}llllll}
    \hline\hline
  $f_i$ &    Frequency (d$^{-1}$)  &   Amplitude (mmag)  & S/N     & Comment \\ \hline
   1   &  14.221393 $\pm$0.000001 & 162.355$\pm$0.135   & 2064.8  & F0, "BM"      \\
   2   &  18.337292 $\pm$0.000001 &  71.842$\pm$0.109   & 1127.4  & F1, "BM"      \\
   3   &   4.115901 $\pm$0.000001 &  53.456$\pm$0.082   & 1114.3  & F1 $-$ F0, "BM"   \\
   4   &  20.496218 $\pm$0.000001 &  27.095$\pm$0.168   &  276.8  & alias (=2$f_{N}$ $-$ 2F0), "BM"  \\
   5   &  10.105493 $\pm$0.000001 &  20.602$\pm$0.055   &  639.0  & 2F0$-$F1, "BM"  \\
   6   &  16.380317 $\pm$0.000001 &  18.114$\pm$0.126   &  247.1  & alias (=2$f_{N}$ $-$ F0 $-$ F1), "BM"   \\
   7   &  22.453196 $\pm$0.000001 &   7.441$\pm$0.054   &  237.8  & 2F1 $-$ F0, "BM"  \\
   8   &  24.326888 $\pm$0.000001 &   5.671$\pm$0.043   &  226.9  & 3F0 $-$ F1, "BM"  \\
   9   &  12.264462 $\pm$0.000002 &   3.945$\pm$0.065   &  103.3  & alias (=2$f_{N}$ $-$ 2F1), "BM"   \\
   10  &  12.062420 $\pm$0.000003 &   3.157$\pm$0.058   &   92.8  & alias (=$f_{8}$ $-$ $f_{9}$), "BM"   \\
   11  &   8.231795 $\pm$0.000003 &   2.753$\pm$0.052   &   90.0  & 2F1$-$2F0, "BM" \\
   12  &  16.178326 $\pm$0.000003 &   2.356$\pm$0.058   &   69.4  & alias (=$f_{6}$ $-$ $f_{9}$ + $f_{10}$), "BM"   \\
   13  &   6.274869 $\pm$0.000004 &   2.065$\pm$0.059   &  59.8   & alias (=$f_{6}$ $-$ $f_{5}$)   \\
   14  &   5.989558 $\pm$0.000004 &   1.872$\pm$0.051   &  62.3   & 3F0 $-$ 2F1 \\
   15  &  14.540628 $\pm$0.000003 &   1.790$\pm$0.043   &  70.5   & $f_1$+$f_m$  \\
   16  &   7.946523 $\pm$0.000004 &   1.657$\pm$0.049   &  57.9   & alias (=2$f_{N}$ $-$ 3F1 + F0 $-$ $f_{9}$ + $f_{10}$) \\
   17  &   2.158970 $\pm$0.000006 &   1.617$\pm$0.068   &  40.7   & alias (=2$f_{N}$ $-$ 2F0 $-$ F1)  \\
   18  &   1.956943 $\pm$0.000006 &   1.415$\pm$0.066   &  36.9   & alias (=2$f_{N}$ $-$ 2F0 $-$ F1 $-$ $f_{9}$ + $f_{10}$)  \\
   19  &  20.211002 $\pm$0.000006 &   1.226$\pm$0.049   &  42.9   & 4F0$-$2F1 \\
   20  &  20.294233 $\pm$0.000007 &   0.935$\pm$0.047   &  33.8   & alias (=2$f_{N}$ $-$ 2F0 $-$ $f_{9}$ + $f_{10}$)   \\
   21  &  22.167916 $\pm$0.000007 &   0.907$\pm$0.045   &  34.8   & alias (=2$f_{N}$ $-$ 3F1 + 2F0 $-$ $f_{9}$ + $f_{10}$) \\
   22  &  22.655153 $\pm$0.000007 &   0.813$\pm$0.040   &  34.5   & alias (=$f_{7}$ + $f_{9}$ $-$ $f_{10}$) \\
   23  &  10.390786 $\pm$0.000007 &   0.767$\pm$0.042   &  31.5   & alias (=2$f_{N}$ $-$ 4F0 + F1) \\
   24  &  22.369961 $\pm$0.000010 &   0.629$\pm$0.044   &  24.3   & alias (=2$f_{N}$ $-$ 3F1 + 2F0) \\
   25  &  16.095095 $\pm$0.000011 &   0.605$\pm$0.047   &  22.1   & 5F0$-$3F1 \\
   26  &  20.177013 $\pm$0.000010 &   0.581$\pm$0.044   &  22.6   & alias (=2$f_{N}$ $-$ 2F0 $-$ $f_m$ )  \\
   27  &   6.072808 $\pm$0.000014 &   0.550$\pm$0.056   &  16.8   & alias (=2$f_{N}$ $-$ 3F0 $-$ $f_{9}$ + $f_{10}$)  \\
   28  &  13.902139 $\pm$0.000017 &   0.380$\pm$0.049   &  13.4   & $f_1$$-$$f_m$  \\
   29  &   1.873768 $\pm$0.000025 &   0.352$\pm$0.065   &   9.3   & 4F0$-$3F1 \\
   30  &   8.148573 $\pm$0.000018 &   0.342$\pm$0.046   &  12.8   & alias (=2$f_{N}$ $-$ 3F1 + F0) \\
   31  &  10.424738 $\pm$0.000017 &   0.311$\pm$0.040   &  13.3   & $f_5$ + $f_m$  \\
   32  &  18.018076 $\pm$0.000018 &   0.307$\pm$0.041   &  12.8   & $f_2$ $-$ $f_m$  \\
   33  &  12.347714 $\pm$0.000019 &   0.305$\pm$0.042   &  12.3   & 3F1 $-$ 3F0 \\
   34  &   0.319314 $\pm$0.000038 &   0.264$\pm$0.073   &   6.2   & $f_m$        \\
   35  &  18.656538 $\pm$0.000023 &   0.239$\pm$0.041   &  10.1   & $f_2$+$f_m$  \\    
\hline
\end{tabular}
\label{tab:Frequency-LC}
\end{center}
\vspace*{-3em}
\tablecomments{The twelve frequencies marked with "BM" in last column are consistent with that in \citet{Bowman2016}. $f_{13}$ to $f_{35}$ are the new detections of this work. Identifications for all of the alias frequencies are listed in parentheses.}
\end{table*}

\begin{table*}
\begin{center}
 \caption{All of the Frequencies Detected in the SC Data (Denoted by $f_{Si}$)}
  \begin{tabular}{@{}lllll}
    \hline\hline
   $f_{Si}$ & Frequency (d$^{-1}$)   & Amplitude (mmag)  &  S/N  & Comment  \\ \hline
    1    &  14.221367 $\pm$0.000015 & 193.985$\pm$0.456  & 728.1 & F0, "BM"         \\
    2    &  18.337228 $\pm$0.000023 &  91.082$\pm$0.326  & 478.4 & F1, "BM"         \\
    3    &  28.442745 $\pm$0.000030 &  65.189$\pm$0.307  & 363.2 & 2F0        \\
    4    &  32.558603 $\pm$0.000023 &  55.284$\pm$0.200  & 474.2 & F0+F1     \\
    5    &   4.115862 $\pm$0.000017 &  50.105$\pm$0.135  & 635.4 & F1$-$F0, "BM"      \\
    6    &  46.779924 $\pm$0.000040 &  27.934$\pm$0.172  & 277.8 & 2F0+F1    \\
    7    &  42.664084 $\pm$0.000044 &  23.076$\pm$0.158  & 250.7 & 3F0       \\
    8    &  10.105444 $\pm$0.000028 &  19.427$\pm$0.086  & 389.0 & 2F0$-$F1, "BM"     \\
    9    &  36.674456 $\pm$0.000054 &  16.834$\pm$0.141  & 204.9 & 2F1       \\
    10   &  22.453101 $\pm$0.000135 &   9.310$\pm$0.195  &  81.7 & 2F1$-$F0, "BM"     \\
    11   &  24.326917 $\pm$0.000163 &   7.705$\pm$0.195  &  67.7 & 3F0$-$F1, "BM"     \\
    12   &  38.548269 $\pm$0.000218 &   3.316$\pm$0.112  &  50.5 & 4F0$-$F1    \\
    13   &   5.990001 $\pm$0.000238 &   2.271$\pm$0.084  &  46.2 & 3F0$-$2F1   \\
    14   &  14.540689$\pm$0.000338  &   2.033$\pm$0.107  &  32.6 & $f_{S1}$+$f_m$   \\
    15   &   8.231774 $\pm$0.000258 &   1.900$\pm$0.076  &  42.6 & 2F1$-$2F0, "BM"    \\
    16   &  40.789789 $\pm$0.000364 &   1.764$\pm$0.100  &  30.2  & 3F1$-$F0    \\
    17   &  20.211252 $\pm$0.000512 &   1.589$\pm$0.127  &  21.5 & 4F0$-$2F1    \\
    18   &  28.761899$\pm$0.000595  &   1.325$\pm$0.123  &  18.5 & $f_{S3}$+$f_m$     \\
    19   &  21.495412$\pm$0.000509  &   1.188$\pm$0.094  &  21.6 & $f_{S10}$$-$3$f_m$ \\
    20   &  42.985775$\pm$0.001171  &   0.742$\pm$0.135  &   9.4 & $f_{S7}$+$f_m$     \\
    21   &  35.717295$\pm$0.000714  &   0.724$\pm$0.081  &  15.4 & $f_{S9}$$-$3$f_m$  \\
    22   &  47.100056$\pm$0.001158  &   0.686$\pm$0.124  &   9.5 & $f_{S6}$+$f_m$    \\
    23   &  32.876915$\pm$0.000909  &   0.628$\pm$0.089  &  12.1 & $f_{S4}$+$f_m$    \\ 
    24   &  34.432726 $\pm$0.001070 &   0.561$\pm$0.093  &  10.3 & 5F0$-$2F1   \\
    25   &  39.833440$\pm$0.001279  &   0.498$\pm$0.100  &   8.6 & $f_{S16}$$-$3$f_m$ \\
    26   &  12.348153 $\pm$0.000827 &   0.497$\pm$0.064  &  13.3 & 3F1$-$3F0   \\  
    27   &  13.906318$\pm$0.001528  &   0.357$\pm$0.085  &   7.2 & $f_{S1}$$-$$f_m$    \\
    28   &  18.024443$\pm$0.002341  &   0.291$\pm$0.106  &   4.7 & $f_{S2}$$-$$f_m$   \\
    29   &  10.419275$\pm$0.002038  &   0.230$\pm$0.073  &   5.4 & $f_{S8}$+$f_m$      \\ \hline\hline
  \end{tabular}
  \vspace*{-3em}
\tablecomments{The frequencies marked with "BM" are consistent with that in \citet{Bowman2016}. Eleven modulation terms modulated by $f_m$ are also annotated in the last column.}
  \label{tab:Frequency-SC}
\end{center}
\end{table*}

\section{DISCUSSION}

To study the amplitude variations of $\delta$ Sct stars, \citet{Bowman2016} extracted the 12 strongest frequencies of the light curves for all $Kepler$ targets with 6400 $\leqslant$ $T_{\rm{eff}}$ $\leqslant$ 10,000 K in KIC using the LC data, and they tracked the amplitude and phase variations of each star observed for four years. That is a suitable choice for stars that pulsate in only a few modes and for most of low-amplitude $\delta$ Sct stars. However, it is possible that some lower amplitude variations still exist after extracting 12 peaks for the HADS stars considering the unprecedented photometric precision of time series data provided by the $Kepler$ $Space$ $Telescope$. A total of 35 significant frequencies are detected with Fourier analysis of the LC data for KIC 5950759 in our work. Apart from the fundamental mode (F0) and the first overtone mode (F1) frequencies, we detect the third independent frequency $f_{m}$ (=0.3193 d$^{-1}$) in the LC data, and the modulation terms modulated by $f_{m}$ in both the LC and SC spectrum. We should note that $f_{m}$ and the modulation terms are not detected by \citet{Bowman2016}, mainly because $f_{m}$ is clearly out of the frequency range of their choosing, and the amplitudes of the modulation terms are weaker than that of the 12 strongest frequencies they extracted.

\subsection{The modulations of F0 and F1}

\begin{figure}
\begin{center}
  \includegraphics[width=0.9\textwidth]{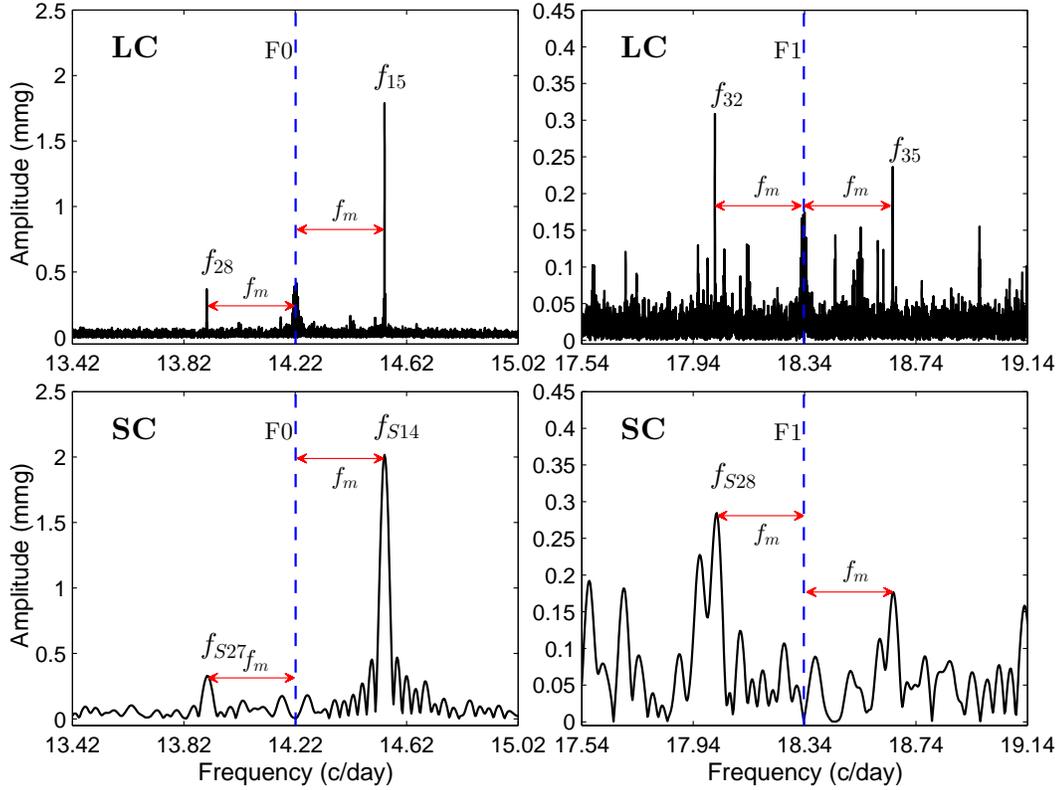}
  \caption{Amplitude spectra after subtracting the main frequencies and its harmonics. The vertical dashed lines indicate the locations of main frequencies F0 and F1. The upper two panels clearly show two pairs of side peaks ($f_{28}$, $f_{15}$ and $f_{32}$, $f_{35}$) around the main frequency F0 and F1 in the LC spectrum. The bottom left panel shows two side peaks around F0 ($f_{S27}$, $f_{S14}$) in the SC spectrum. The bottom right panel also shows two side peaks around F1 in the SC spectrum, however, the S/N of the right side peak around F1 is lower than the significant criterion of this work.}
    \label{fig:modulation1}
\end{center}
\end{figure}

Two groups of side peaks around F0 and F1 (i.e. $f_{28}$ = 13.90214 $\rm{d^{-1}}$ and $f_{15}$ = 14.54063 $\rm{d^{-1}}$, $f_{32}$ = 18.01808 $\rm{d^{-1}}$, and $f_{35}$ = 18.65654 $\rm{d^{-1}}$; see upper two panels in Figure \ref{fig:modulation1}) seem to be the most interesting features in the amplitude spectrum of the LC data of KIC 5950759. The bottom left panel of Figure \ref{fig:modulation1} shows two side peaks around F0 ($f_{S27}$, $f_{S14}$) detected in the SC spectrum. As shown in the bottom right panel of Figure \ref{fig:modulation1}, there are also two side peaks around F1 in the SC spectrum, however, only the left side peak ($f_{S28}$) is detected because its S/N is higher than the significant criterion we adopted. The side peaks around F0 and F1 form two pairs of uniformly spaced triplets with a frequency interval of $f_{m}$ = 0.3193 d$^{-1}$. CoRoT 101155310 was once found to have a periodic modulation ($f_{m}$= 0.193 d$^{-1}$) through $CoRoT$ data by \citet{Poretti2011}, but later this modulation was thought to be an instrumental effect and should be removed in the light curve \citep{Poretti2015}. 
However, several $\delta$ Sct stars observed by $Kepler$ show equally spaced frequency components in their frequency spectra produced by intrinsic variation \citep{Breger2011}. Therefore, the modulation of the main components F0 and F1 of KIC 5950759 could be caused either by a periodic instrumental effect or an intrinsic variation.

\subsubsection{Instrumental effect}

To investigate the triplets, we first check if these side peaks are caused by instrumental effects: the $Kepler$ orbital period is $P_{\rm{orb}}$ = 372.5 days (the corresponding frequency is $f_{\rm{orb}}$ = 0.00268 d$^{-1}$), the $Kepler$ rotates 90$^{\circ}$ every 93 days ($f_{\rm{rot}}$ = 0.011 d$^{-1}$), the process of data downlink creates an interruption (lasting 24 hr) every 32 days ($f_{\rm{downlink}}$ = 0.031 d$^{-1}$), and the momentum desaturation of the reaction wheel for $Kepler$ happens every 2.98 days ($f_{\rm{reaction}}$ = 0.336 d$^{-1}$) \citep{Hass2010,Van Cleve2016}. The Rayleigh frequency resolution of the LC data is 0.00068 d$^{-1}$ since the time base is about 1460 days. Therefore, the side peaks detected in this work are not from the known instrumental effects as no frequency listed above is equal to $f_{m}$ within the Rayleigh frequency resolution. Although the possibility is rather small, the triplet structures in the spectra could still be produced by some unknown instrumental effects.

\subsubsection{A new radial mode or non-radial mode?}

The HADS stars are usually mono-mode or double-mode radial variables \citep{Breger2000}. Improving of the detection ability, low-amplitude radial and nonradial modes are also detected in some HADS stars that are previously considered to be pure mono-mode or double-mode radial variables (e.g. AI Vel, \citealt{Walraven1992}; DY Peg, \citealt{Garrido1996}). The multiplet structures are often shown in the frequency spectra of the above HADS stars when the long time and high-precision photometric data are obtained. For the exact equidistant frequency triplet structures in $\delta$ Scuti stars, \cite{Breger2006} provide an explanation named the "Combination Mode Hypothesis,." In this hypothesis, the highest amplitude mode $f'_{1}$ and a real second mode $f'_{2}$ are excited, and the harmonic of $f'_{1}$ (e.g. 2$f'_{1}$) is also likely to occur in the spectrum, then the combination  2$f'_{1}$ - $f'_{2}$ may also occur. This combination is observed at $f'_{3}$ and a frequency triplet (i.e. $f'_{3}$, $f'_{1}$, and $f'_{2}$) is observed around $f'_{1}$.

Under this hypothesis, the small amplitude frequency $f_{15}$ in the LC spectrum of KIC 5950759 is suspected of a new mode, and $f_{28}$ is thought to be the combination of 2F0 and $f_{15}$ (i.e. 2F0 - $f_{15}$). Other frequencies (apart from the combinations and harmonics of F0 and F1) can also be combinations involving F0, F1, and $f_{15}$. Thus, this star might be a new triple-mode variable. However, the period ratio P1/P0 (=0.7755)(the first overtone and the fundamental mode) and amplitude ratio $A_{f_{15}}$/$A_{F}$ (=0.011) are inconsistent with that of all the known triple-mode radial variables \citep{Wils2008}. For this reason, $f_{15}$ may not be a new radial mode in KIC 5950759. 

If $f_{15}$ is considered to be a nonradial mode, the equidistant triplets in frequency spectra can occur only when stars rotate extremely slowly. In the case of stellar rotation exceeding a few km s$^{-1}$, the triplets caused by rotation will not be equally spaced. Hence, $f_{15}$ is unlikely a nonradial mode in KIC 5950759.

\subsubsection{Blazhko effect?}

Equidistant triplets structures are often shown in the Fourier spectra of the Blazhko RR Lyrae stars and the frequency separation of the triplets is identical with the modulation frequency \citep[e.g.][]{Smith1999,Jurcsik2005,Kolenberg2006}. In Blazhko RR Lyrae stars, the Blazhko modulation frequency can also be directly detected in the spectrum with long baseline and high-precision observations. For KIC 5950759, the equidistant triplet structures in the LC spectrum are similar to that in Blazhko RR Lyr stars, and a modulation frequency is also clearly detected in the LC spectrum. These features imply that the triplets in KIC 5950759 are related to the Blazhko effect. However, we should note that there is no obvious modulation feature of the amplitude in the light curve of KIC 5950759 as shown in Blazhko RR Lyrae stars. Moreover, the light curve of Blazhko RR Lyrae stars (e.g. CoRoT 101128793; \citealt{Poretti2010}) shows bumps and larger scatter at a well-defined phase after subtracting all the significant pulsation frequencies (see Fig.3 in \citealt{Poretti2010}). In the case of KIC 5950759, the residuals after subtracting all of the pulsation frequencies do not show similar bumps and scatter as in Blazhko RR Lyrae stars. Therefore, there is no sufficient evidence so far to support the possibility that the triplets are caused by the Blazhko effect observed in RR Lyrae stars.

\subsubsection{Amplitude modulation with rotation?}

The $Kepler$ mission has found several $\delta$ Sct stars in which their frequency spectra show uniformly spaced multiplets caused by small modulation of the amplitudes with rotation  \citep{Breger2011}. A small amplitude modulation with a frequency of 0.1597 d$^{-1}$ was seen in $\delta$ Sct star KIC 9700322, and this was interpreted to be the rotation frequency of the star \citep{Breger2011}. For KIC 5950759, the equally spaced triplets (frequency interval $f_{m}$ = 0.3193 d$^{-1}$) in its frequency spectra are most likely originated from  modulation of the amplitudes with rotation. Mathematically, we can describe the effect of amplitude modulation as follows: considering a simple case in which a single periodic signal is modulated with a general periodic function \citep{Benko2011}. For a double-mode HADS star, its pulsation can be expressed as

\begin{equation}
 S_{pul} = \sum_{i=1}^{N_{pul}}A_{i}sin[2\pi f_{pul}it + \phi_{i}^{pul}]
\end{equation}

where $A_{i}$, $f_{pul}i$, and $\phi_{i}^{pul}$ are the amplitude, pulsation frequency, and phase, and $N_{\rm{pul}}$ is 2. The modulation function can be expressed as

\begin{equation}
 S_{mod} = \sum_{j=1}^{N_{mod}}B_{j}sin[2\pi f_{mod}jt + \phi_{j}^{mod}]
\end{equation}

where $B_{j}$, $f_{mod}j$, and $\phi_{j}^{mod}$ are the amplitude, modulation frequency, and phase, and $N_{mod}$ is 1. Then the pulsation signal with modulation function can be written as

\begin{equation}
S_{Total} = \sum_{i=1}^{N_{pul}}A_{i}sin[2\pi f_{pul}it + \phi_{i}^{pul}]\sum_{j=1}^{N_{mod}}B_{j}sin[2\pi f_{mod}jt + \phi_{j}^{mod}]
\end{equation}

Using Simpsons rule, $S_{\rm{Total}}$ can be re-written as a sum,

\begin{equation}
\begin{split}
S_{Total} = \sum_{i=1}^{N_{pul}}\sum_{j=1}^{N_{mod}}\frac{A_{i}B_{j}}{2}(cos[2\pi (f_{pul}i - f_{mod}j)t + \phi_{i}^{pul} - \phi_{j}^{mod}] \\
- (cos[2\pi (f_{pul}i + f_{mod}j)t + \phi_{i}^{pul} + \phi_{j}^{mod}])
\end{split}
\end{equation}

Thus, if the modulation frequency is considered to be rotation frequency, we can see frequency peaks produced by the linear combinations of the pulsation frequency and the rotation frequency in the Fourier transform of the above $S_{\rm{Total}}$. For KIC 5950759, these two pairs of side peaks in the amplitude spectrum can be considered as the modulation of its main pulsation modes with rotation frequency $f_{m}$ = 0.3193 $\rm{d}^{-1} $ of the star. We note that an accurate rotation velocity from high-resolution spectroscopic observations will help us verify the fact that the modulation is indeed from stellar rotation. Given that the modulation terms of KIC 5950759 was caused by stellar rotation, considering one of the two HADS stars (the other one is KIC 9408694) observed by $Kepler$ is detected with this weak modulation effect, it seems that this kind of modulation effect may not be a rare phenomenon in HADS stars. However, the amplitude ratios of the fundamental frequency (F0) and the side peaks ($f_{15}$ and $f_{28}$) are about 90 $\sim$ 400. Therefore, it is not easy to observe such weak modulation terms from ground-based telescopes as the amplitudes of these terms are most possibly lower than the detected threshold of the telescopes.

\subsection{The location in the H-R diagram}

For KIC 5950759, the ratio of F0 = 14.221373 d$^{-1}$ and F1 = 18.337249 d$^{-1}$ gives 0.7755, which is in the range of typical period ratio for the double-mode HADS stars. With about 153 $\delta$ Scuti and SX Phoenicis stars, \cite{Poretti2008} provided the period-luminosity relationship as
$M_{V} = -1.83(\pm0.08) - 3.65(\pm0.07)$ $\rm{log}$ $P_{F}$, where $P_{F}$ is the period of the fundamental mode. We get $M_{V} = 2.38(\pm0.16)$ mag for KIC 5950759 using $\rm{log}$ $P_{F} = -1.153$.

\begin{figure}
\begin{center}
  \includegraphics[width=0.90\textwidth]{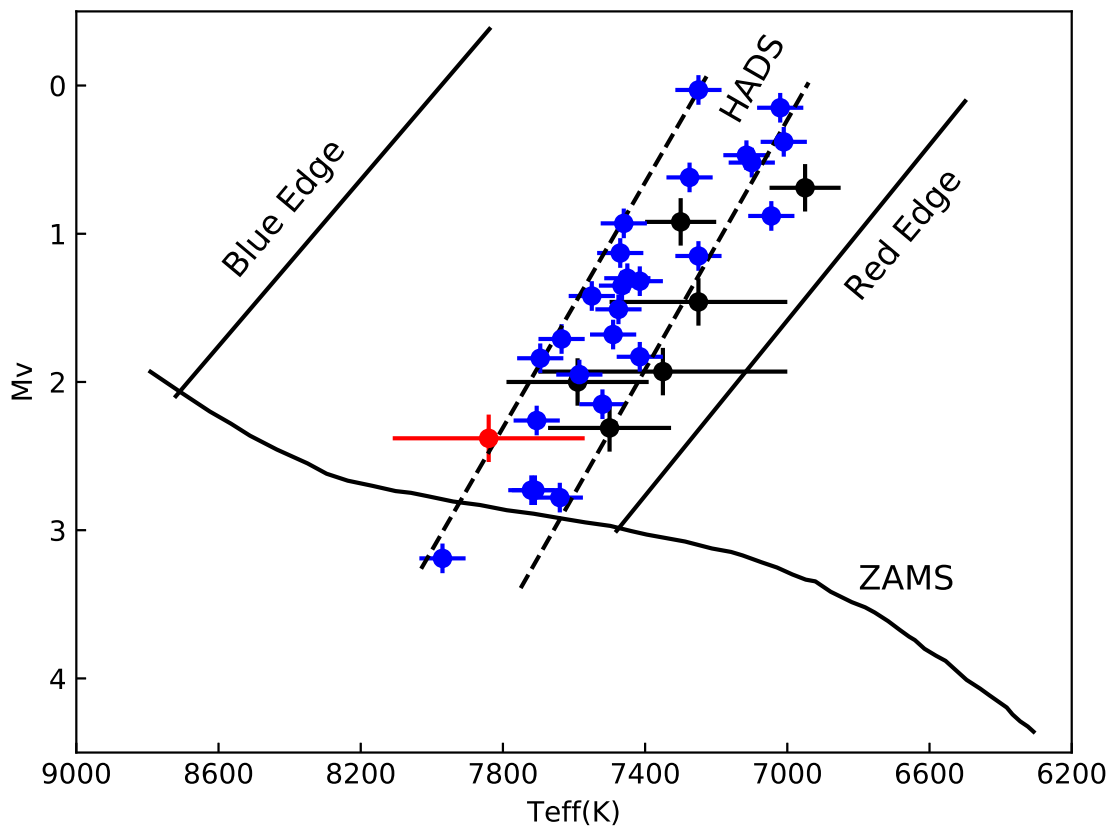}
  \caption{Location of 32 HADS stars and KIC 5950759 in the H-R diagram. KIC 5950759 is shown as a red dot. The blue dots are stars from table 2 of \citet{McNamara2000}. The black dots are stars collected in the literature \citep{Poretti2005,Poretti2011,Christiansen2007,Balona2012a,Ulusoy2013,Pena2016}. The zero-age main sequence (ZAMS), HADS instability strip (dashed lines), and $\delta$ Scuti instability strip (solid lines) are from \cite{McNamara2000}.}
    \label{fig:HR_IS}
\end{center}
\end{figure}

Based on the above $M_{V} = 2.38(\pm0.16)$ mag and $T_{\rm{eff}} = 7840(\pm300)$ K (listed in KIC), the location of KIC 5950759 in the H-R diagram is shown in Figure \ref{fig:HR_IS}. An additional 32 HADS stars collected from the  literature \citep{McNamara2000,Poretti2005,Poretti2011,Christiansen2007,Balona2012a,Ulusoy2013,Pena2016} are listed in Table \ref{tab:parameters} and are also shown in Figure \ref{fig:HR_IS}. In this figure, KIC 5950759 is in the bottom of the HADS instability strip and likely situated in the main sequence. However, \citet{Huber2014} note that the KIC temperatures for stars with $T_{\rm{eff}}$ $>$ 6500 K are on average 200 K lower than that obtained from Sloan photometry or the infrared flux method \citep{Pinsonneault2012}, and provide $T_{\rm{eff}} = 8040(\pm270)$ K for this star, which puts it significantly away from the HADS instability strip, indicating the possibility of this star being outside of the instability strip. We should note that the location of KIC 5950759 in the H-R diagram cannot yet be determined well due to the large uncertainty on its effective temperature. High-resolution spectroscopic observations are needed to obtain a more accurate effective temperature, which would provide a relatively tight constraint on the location of KIC 5950759 in the H-R diagram.

\begin{table*}
\begin{center}
\caption{Parameters of 32 HADS Stars}
  \begin{tabular}{@{}llllllllll}
    \hline\hline
    ID & Star &  log $P$  &  $T_{\rm{eff}}$ &  $M_{\rm{V}}$  &ID & Star & log $P$ & $T_{\rm{eff}}$ &  $M_{\rm{V}}$  \\
       &           &             &    (K)          & (mag) &       &        &      &   (K) &    (mag)   \\\hline
    1  & BL Cam    &   -1.408    &   7970  &  3.19 & 17 &  RS Gru    &   -0.833    &   7470  &  1.13    \\
    2  & SX Phe    &   -1.260    &   7710  &  2.73 & 18 &  DY Her    &   -0.828    &   7250  &  1.15    \\
    3  & KZ Hya    &   -1.225    &   7640  &  2.78 & 19 &  V567 Oph  &   -0.825    &   7460  &  0.93    \\
    4  & CY Aqr    &   -1.214    &   7720  &  2.73 & 20 &  VZ Cnc    &   -0.749    &   7045  &  0.88   \\
    5  & DY Peg    &   -1.137    &   7705  &  2.26 & 21 &  BS Aqr    &   -0.740    &   7275  &  0.62   \\
    6  & GP And    &   -1.104    &   7520  &  2.15 & 22 &  VX Hya    &   -0.651    &   7100  &  0.52   \\
    7  & AE UMa    &   -1.065    &   7585  &  1.95 & 23 &  RY Lep    &   -0.647    &   7115  &  0.47   \\
    8  & EH Lib    &   -1.054    &   7695  &  1.84 & 24 &  DE Lac    &   -0.596    &   7010  &  0.38   \\
    9  & RV Ari    &   -1.031    &   7415  &  1.83 & 25 &  V1719 Cyr &  -0.573     &   7020  &  0.15   \\
    10 & BE Lyn    &   -1.018    &   7635  &  1.71 & 26 &  SS Psc    &   -0.541    &   7250  &  0.03   \\
    11 & YZ Boo    &   -0.983    &   7490  &  1.68 & 27 &  GSC 00144-03031 &  -1.234  & 7590$\pm$200  & 2.00   \\
    12 & BP Peg    &   -0.960    &   7550  &  1.42 & 28 &  UNSW-V-500      &  -1.134  & 7500$\pm$173  & 2.31   \\
    13 & AI Vel    &   -0.952    &   7475  &  1.51 & 29 &  BO Lyn          &  -1.030  & 7350$\pm$350  & 1.93    \\
    14 & SZ Lyn    &   -0.919    &   7465  &  1.35 & 30 &  CoRoT 101155310 &  -0.900  & 7250$\pm$250  & 1.46    \\
    15 & AD CMi    &   -0.910    &   7450  &  1.30 & 31 &  KIC 9408694     &  -0.753  & 7300$\pm$150  & 0.92    \\
    16 & XX Cyg    &   -0.870    &   7415  &  1.32 & 32 &  KIC 6382916     &  -0.691  & 6950$\pm$100  & 0.69    \\\hline
  \end{tabular}
  \vspace*{-3em}
\tablecomments{Stars with ID 1-26 are from \cite{McNamara2000}, the corresponding errors in $T_{\rm{eff}}$ and $M_{\rm{V}}$ are 65 K and 0.1 mag. Stars with ID 27-32 are from \cite{Poretti2005}, \cite{Christiansen2007}, \cite{Pena2016}, \cite{Poretti2011}, \cite{Balona2012a}, \cite{Ulusoy2013}, respectively. $M_{V}$ is calculated using the period-luminosity relationship provided by \citet{Poretti2008}.}
  \label{tab:parameters}
\end{center}
\end{table*}

\section{SUMMARY}

Based on four years of uninterrupted time series photometric data from the $Kepler$ $Space$ $Telescope$, we analyze the pulsations of KIC 5950759 in-depth, and extract 35 and 29 significant frequencies from the LC and SC data, respectively, including the fundamental (F0) and first overtone mode (F1) frequencies, their combinations, harmonics, and the modulation terms.

Apart from F0 and F1, the third independent frequency ($f_{m}$ = 0.3193 d$^{-1}$) and a small amplitude modulation of $f_{m}$ to the two independent modes are detected in the light curves of KIC 5950759. The modulation frequency $f_{m}$ is detected in LC with an S/N of 6.2. The $f_{m}$ is not detected in the SC data most possibly because of the lower data precision of the SC observations than that of LC and the shorter observing duration. This is the first detection of the modulation effect in a double-mode HADS star. The modulation frequency is not equal to any frequency caused by the known instrumental effects of $Kepler$. We discussed three potential explanations, i.e. a new radial mode or nonradial mode, the Blazhko effect, and amplitude modulation with rotation for the equidistant triplets structures in frequency spectra. The most possible cause for the triplets is amplitude modulation with the star's rotation frequency as 0.3193 $\rm{d}^{-1} $.

With frequency of the fundamental radial mode, the absolute visual magnitude for KIC 5950759 is calculated as $M_{V} = 2.38 \pm 0.16 $ mag. The parameters ($M_{V}$ and $T_{\rm{eff}}$ = 7840$\pm300$ K) suggest that KIC 5950759 is in the bottom of the HADS instability strip and likely situated in the main sequence in the H-R diagram. Spectroscopic observations and multicolor photometric time series are necessary to deepen our understanding to the modulation terms and evolutionary stage of the star.

\acknowledgements

The authors thank the referee for the very helpful comments. This research is supported by the program of the light in China's Western Region (LCWR; grant Nos. XBBS-2014-25, 2015-XBQN-A-02), the National Natural  Science Foundation of China (grant Nos.11273051, 11661161016), the 13th Five-year Informatization Plan of Chinese Academy of Sciences (grant No. XXH13503-03-107), the Youth Innovation Promotion Association CAS (grant Nos. 2014050, 2018080), Heaven Lake Hundred-Talent Program of Xinjiang Uygur Autonomous Region of China, and the Strategic Priority Research Program of the Chinese Academy of Sciences, Grant No. XDB23040100. We wish to thank the $Kepler$ science team for providing such excellent data.




\clearpage

\end{document}